\documentclass[journal]{IEEEtran}
\usepackage[utf8]{inputenc}
\usepackage{graphicx}
\usepackage{subfigure}
\usepackage{textcomp}
\usepackage{amsmath}
\usepackage{amsfonts}
\usepackage{xspace}
\usepackage{cite}
\usepackage{url}
\usepackage{enumerate}
\usepackage{epstopdf}
\usepackage{color}
\usepackage{color,soul}

\usepackage[]{todonotes}

\graphicspath{{figures/},{figurepdfs/}}
\newcommand{\Lower}[1]{\smash{\lower 1.5ex \hbox{#1}}}

\def\putbox#1#2#3#4{\makebox[0in][l]{\makebox[#1][l]{}\raisebox{\baselineskip}[0in][0in]{\raisebox{#2}[0in][0in]{\scalebox{#3}{#4}}}}}
\def\rightbox#1{\makebox[0in][r]{#1}}
\def\centbox#1{\makebox[0in]{#1}}
\def\topbox#1{\raisebox{-0.60\baselineskip}[0in][0in]{#1}}
\def\midbox#1{\raisebox{-0.20\baselineskip}[0in][0in]{#1}}

\usepackage{array}
\newcolumntype{P}[1]{>{\centering\arraybackslash}p{#1}}
\newcolumntype{M}[1]{>{\centering\arraybackslash}m{#1}}   

\begin{document}
\title{Low-Frequency Noise Reduction Using In-Pixel Chopping To Enhance The Dynamic Range of a CMOS Image Sensor.}
\author{Kapil Jainwal, \textit{Member IEEE}
\thanks{Manuscript received June X, XXXX.}
\thanks{Author is with the Electrical Engineering Department, Indian Institute of Technology Delhi, 110016 New Delhi, India. (e-mail: KJ: kapiljainwal@gmail.com) 
}
}
\markboth{Journal of \LaTeX\ Class Files,~Vol.~13, No.~9, January~2016}%
{Shell \MakeLowercase{\textit{et al.}}: Bare Demo of IEEEtran.cls for Journals}
      
\maketitle
\begin{abstract}
\label{sec:ABSTRCT}
{\color{black}In this paper, a low-noise CMOS image sensor with enhanced dynamic range (DR), using an in-pixel chopping technique, is presented. The proposed in-pixel chopping technique is used to reduce the low-frequency or 1/\textit{f} noise of the source follower (SF) in an active pixel sensor (APS), which is a major component of the temporal noise. A conventional 3T active pixel, with n-well/p-sub photodiode (PD), is modified to implement a chopper inside a pixel. A single minimum sized nMOS transistor is used in each pixel, without much compromising in the fill-factor (FF). Using chopping action the low-frequency noise of the source follower is modulated to the chopping frequency (\textit{$\textbf{f}_\textbf{ch}$}) which is much higher than the maximum frequency of the input signal frequency band. The up-converted low-frequency noise is eliminated using a column level low-pass filter (LPF), in the later stage. The reduction in the temporal noise also results in an enhanced dynamic range {\color{black}} of the image sensor. In addition, the readout consists of a column level high gain chopper amplifier also reduces the non-linearity of the source follower.  
	To validate the proposed technqiue a prototype sensor, consists of a 128$\times$128 sized pixel array with in-pixel chopping and column level read-out circuitry, is fabricated in AMS 0.35 \textbf{$\mu$}m CMOS OPTO process. The pixel pitch is 10.5 \textbf{\textbf{$\mu$}}m (horizontal and vertical both) with a fill-factor of around 30$\textbf{\%}$. The temporal noise is measured as 280 $\mu$V$_{\text{\textbf{rms}}}$ at the chopping frequency (\textit{$\textbf{f}_\textbf{\textbf{ch}}$}) of 8 MHz,  which shows a reduction in the noise power by 11 dB. Due to reduced noise floor the dynamic range is enhanced from 65 dB to 76 dB, using the proposed technique. }

\end{abstract}
%
\begin{IEEEkeywords}
Low-frequency noise, 1/\textit{f} noise, chopper amplifier, CMOS image sensors, dynamic range.
\end{IEEEkeywords}
%
\section{Introduction}
\label{sec:INTRO}
\IEEEPARstart In the recent development of digital imaging systems, CMOS image sensors have replaced charged-coupled-devices (CCDs) in many fields, due to low-power consumption, easy on-chip integration of transistors and photo-sensors, low-cost fabrication and ease to implement complex functionality. However, the high noise is a major bottleneck in the imaging performance of a CMOS image sensor. High dynamic range (DR), which is one of the primary performance defining parameters for a CMOS image sensor, is limited by a limited output swing and high noise.
The primary sources of noise in an active pixel sensor (APS) of a CMOS imager are the thermal noise from the switches and the low-frequency $1/f$ noise from the source follower (SF). The thermal noise from the reset switch of the pixel can efficiently be reduced using correlated double sampling (CDS) \cite{paper:KANSY_CDS,paper:CDS86,paper:wangTHESIS,paper:Enz_CDS_chop} . The low-frequency or 1/$f$ noise of the SF remains as a major source of noise in an active pixel.
The $1/f$ noise is usually caused by random trapping and detrapping of charge carriers into the unsaturated energy states or dangling bonds, present at the gate Si-SiO$_2$ interface. This causes the discrete fluctuation of the conducting current or the threshold voltage. This low-frequency noise phenomenon is temporal in nature and due to blinking behavior, it is very much visible to human eyes. The $1/f$ noise power spectral density (PSD) decreases with increases in the area of a transistor, thus, it severely affects the quality of an image captured from an imager fabricated in sub-micron deeper technology levels. \par
Bloom and Nemirovsky \cite{paper:SWITCH_Bloom91} have introduced the 1/$f$ noise reduction technique based on cycling a MOS transistor between strong inversion and accumulation region. The switching technique is further investigated, modeled and verified by several researchers  \cite{paper:SWITCH_Gierkink99,paper:SWITCH_Klumperink2000,paper:SWITCH_VEN_DER_2000}. This technique is commonly used in many systems like phase locked loop (PLL), in which switching is used to reduce the phase noise of a voltage controlled oscillator (VCO). In \cite{paper:Mypaper1,paper:Mypaper2} the switching of MOS transistor is shown to reduce the $1/f$ noise in APS using shared source followers among the pixels. However, in this technique, the noise reduction is limited by the number of shared source followers.  \par
\begin{figure*}
	\centering
	\includegraphics[scale =1.2 ]{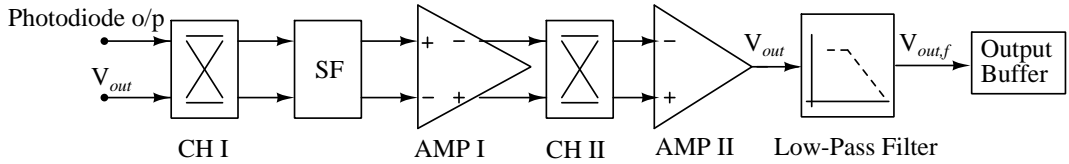}
	\vspace{-0.3 cm}
	\caption{{\color{black}Block diagram for in-pixel chopping.  }}
	\vspace{-0.4 cm}
	\label{fig:BD_INPXL}
\end{figure*}

\par
Recently reported imagers are using correlated multiple sampling  (CMS) with high column-level gain \cite{paper:Lim,paper:Seo} in read-out chain along with  BSF as buffer \cite{ISSCC2012,paper:Chen2,CMS_SJ2012} or p-MOS amplifier \cite{ISSCC2011} which exhibits lower dark random noise and achieved input referred noise below one electron.
In~\cite{TED16} a thin oxide pMOS transistor for pixel-
level buffering of the sense node voltage, instead of conventional thick-oxide nMOS transistor. A thin oxide pMOS transistor exhibits a reduced $1/f$ noise and the sensor obtained a sub-electron level input referred noise.
\par
The use of pMOS transistor reduces the fill-factor of the pixel. A buried channel source follower (BSF) instead of surface-mode nMOS transistor as in-pixel buffer reduces the dark random noise. In BSF the low-frequency noise reduction occurs due to the conduction of the charge carriers away from the Si-SiO$_2$ interface. This prevents the interaction of mobile charge carriers with defect states and eventually decreases the random conduction. The negative threshold voltage of berried channel transistor which helps in the output swing improvement. However, the output swing improvement could lead to the higher temporal noise and image lag. Thus, there is a trade-off between exists between the reduction in noise and output swing enhancement.
A buried channel SF and thin oxide pMOS both techniques required device level modifications and thus, would increase the cost of the sensor.


Auto-zeroing (AZ),~a sampling-based technique, is also used for $1/f$ noise reduction \cite{paper:Enz_CDS_chop,paper:CHOP4}. The AZ works as a high-pass filter and thus, reduces the low-frequency noise. However, due to the sampling action involved in AZ the thermal noise floor (a wide-band noise) increases due to aliasing. Thus, the low-frequency noise reduction comes at the cost of thermal noise. CDS, which is a special case of AZ is used in image sensors for 1/$f$ noise reduction, however, the reduction depends on the sampling frequency of the CDS and is only effective when the sampled noise components are correlated \cite{paper:Enz_CDS_chop,  paper:KANSY_CDS, paper:wangTHESIS,paper:CDS86}. 
\par
Chopping  \cite{MICRO_POWER_CHOP, GRAY_CHOP, paper:CHOP4, WITTE_CHOP,paper:Chop_COLN_81,paper:Enz_CDS_chop,paper:Chop_FOTOUHI_94} is one the most popular techniques used for the $1/f$ noise reduction. The chopping is based on modulation in which the low-frequency noise is up-converted to the chopping frequency ($f_{ch}$) far beyond the frequency band of interest. Chopping needs extra switches and would hamper the fill-factor of the pixel and thus has never been used in a pixel. In this work, a novel technique is presented to implement chopping inside a {\color{black}conventional 3T pixel}. One switch and a signal line per pixel are additionally used to modify conventional 3T pixel, to implement in-pixel chopping, which hampers the fill-factors on the pixel in the least amount.  
%
%

%
The low-frequency noise reduction does not depends on the size of the transistor and thus, a minimum sized SF is used in the pixel, which compensates the increase in the fill-factor due to additional switch.
{\color{black}It is shown later that the low-frequency noise power reduces by around 11 dB, as compared to a conventional 3T APS without chopping (and only employed with double sampling (DS)). Consequently the dynamic range of the imager enhances from 65 dB to 76 dB.
	\par   
	{The rest of the paper is organized as follows:} {\color{black} Imager system including the in-pixel chopping implementation  is described in section \ref{sec:SYS}. In section \ref{sec:CHOP_ANALISYS} analysis of the effect of chopping on thermal and low-frequency noise is explained. Prototype sensor overview along with measurement results are presented in section \ref{MSR_SIM}. Along with a performance comparison of the proposed imager with other recently reported works, the paper is concluded in section \ref{CONCLUSION}}.
%
%

 \begin{figure*}
 	\centering
 	\def\putbox#1#2#3#4{\makebox[0in][l]{\makebox[#1][l]{}\raisebox{\baselineskip}[0in][0in]{\raisebox{#2}[0in][0in]{\scalebox{#3}{#4}}}}}
 	\def\rightbox#1{\makebox[0in][r]{#1}}
 	\def\centbox#1{\makebox[0in]{#1}}
 	\def\topbox#1{\raisebox{-0.60\baselineskip}[0in][0in]{#1}}
 	\def\midbox#1{\raisebox{-0.20\baselineskip}[0in][0in]{#1}}
 	\includegraphics[scale = 1.1]{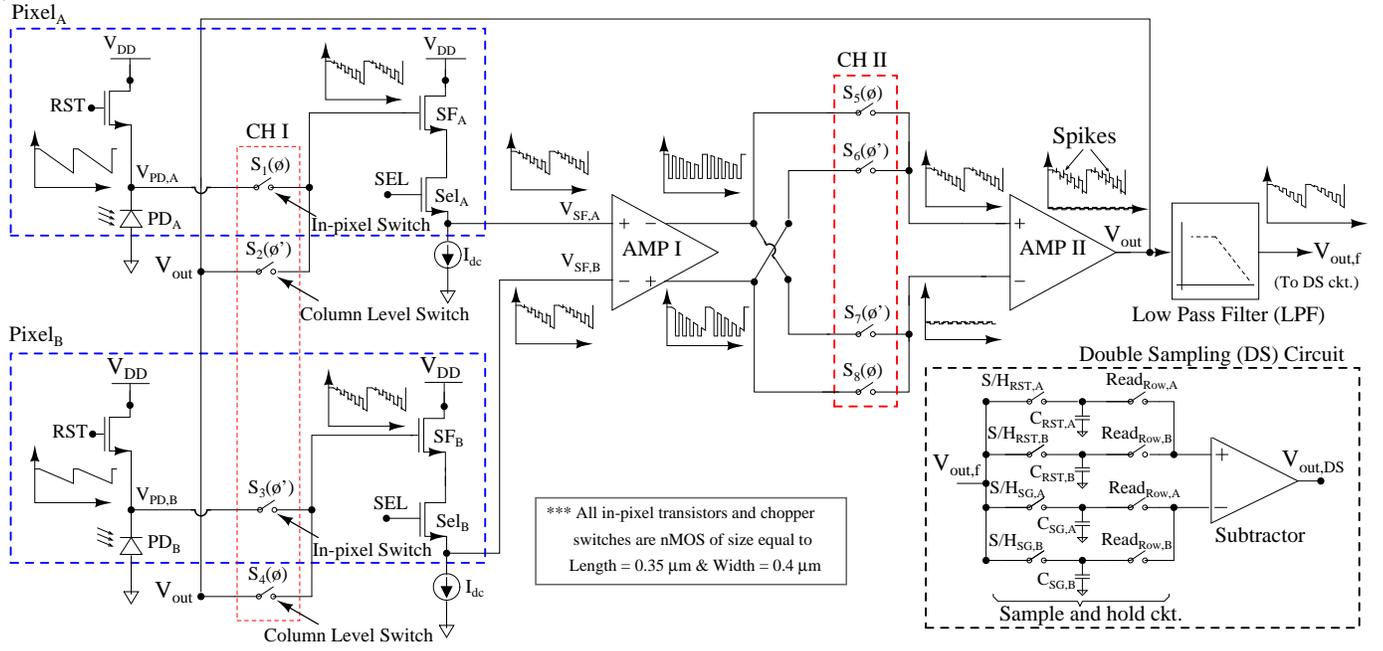}
 	\caption {{\color{black}Circuit diagram of in-pixel chopping - excluding  Pixel{$_A$} and Pixel{$_B$}, other blocks are placed in column level readout circuitry}}
 	\label{Ckt_inPxl_Chop}                                                      \end{figure*}

 \begin{figure}
 	\centering
 	\includegraphics[scale =1 ]{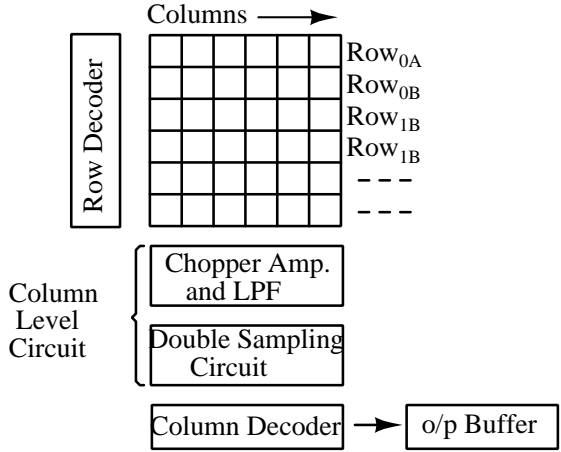}
 	\caption{{\color{black}Image sensor architecture.  }}
 	\label{fig:ARCH} 
 \end{figure}

 \section{System Design}
  \label{sec:SYS}

 \subsection{Block Diagram}
 \label{sec:BD}
The basic building block of the proposed in-pixel chopping is shown in Fig. \ref{fig:BD_INPXL}. The photodiode (PD) output signal is modulated to the chopping frequency ($f_{ch}$) using the first chopper (CH I) before being buffered by the SF. The output of the SF is fed to the input of the amplifier stage I (AMP I). The output of the AMP I is composed of the amplified modulated PD signal, amplified low-frequency noise from the SF, and the noise and output offset of the AMP I. The output of the AMP I is chopped again using the second chopper (CH II). The AMP I and the CH II are placed in the column and does not affect the fill-factor of the pixel. The CH II demodulates the PD signal to its original baseband frequency whereas, modulates the low-frequency noise of the SF and AMP I, and amplifier offset voltage to $f_{ch}$. After CH II the signals are further amplified by the amplifier stage II (AMP II). The output of the AMP II is fed back to the other input of CH I to complete the closed loop unity gain feedback configuration. 
During chopping action ripples are generated in the output due to clock feed-through of the overlapping capacitance present between drain and gate of the switching transistor \cite{paper:Clock_feed_thourgh1}. A low-pass filter (LPF) followed by CH II suppresses the up-modulated offset and 1/$f$ noise and also blocks the spikes or ripples in the output. \cite{paper:Filter1,paper:Filter2} This low-pass filter is  placed in the column of the CMOS image sensor, which does not affect the fill factor of the pixel. 
The modified pixel read-out helps in achieving the functionality as well as a reduction in the low-frequency noise. 

 \subsection{Circuit Design}
 \label{sec:Cir_Disgn}
{\color{black}
 The circuit diagram for the in-pixel chopping in active pixel sensor of a CMOS imager is shown in {\color{black}Fig. \ref{Ckt_inPxl_Chop}}.   
  {\color{black}The two pixels A and B consist of photodiodes PD$_A$, PD$_B$,  chopper switches S$_1$, S$_3$ (of CH I), select switches Sel$_A$, Sel$_B$, and source followers SF$_A$, SF$_B$. 
  The remaining two switches S$_2$, S$_4$ of CH I, AMP I, CH II (switches S$_5$-S$_8$), AMP II, a low-pass filter, and a double sampling circuit are placed in column level read-out circuits. At the onset, the photodiodes of Pixel$_{A}$ and Pixel$_{B}$ are reset to $V_{rst}$ using RST switch. Light is then integrated on the photodiodes. After the integration time, the photodiode output signals V$_{PD,A}$ and V$_{PD,B}$ are modulated to chopping frequency $f_{ch}$ using switches S$_1$-S$_4$ of CH I. The non-overlapping clock signals $\phi$ and ${\phi}$' run at the fundamental chopping frequency $f_{ch}$.  
During readout, Pixel$_{A}$ and Pixel$_{B}$ are selected together using the select signal SEL at the input of switches Sel$_A$ and Sel$_B$. The row decoder of the imager selects two rows at a time for simultaneous selection of two adjacent pixels in the column. The SF output of Pixel$_{A}$ and Pixel$_{B}$ are amplified using AMP I. The AMP I is realized using a folded cascode differential input differential output amplifier.

 The clock signal $\phi$ turns {\color{black}the switches S$_1$, S$_4$} and S$_5$, S$_8$  ON  for a  time interval $t_1$ and the clock signal ${\phi}$' turns {\color{black}the switches S$_2$, S$_3$} and S$_6$, S$_7$ ON, for $t_2$  ($t_1$ and $t_2$ are non-overlapping and equal time intervals). After modulation of the photodiode signals V$_{PD,A}$ and V$_{PD,B}$ through CH I 
  and buffered by the SF, the input of the AMP I can be given as
\begin{equation}
\begin{aligned}
V_{SF,A} = V_{PD,A}(t_1) + V_{out}(t_2) + N_{sf,A} , \\
V_{SF,B} = V_{PD,B}(t_2) + V_{out}(t_1) + N_{sf,B} ,
\end{aligned}
\end{equation}

where N$_{sf,A}$ and N$_{sf,B}$ are the low-frequency noise from SF$_A$ and SF$_B$, respectively. The notation of V$_{PD,A}(t_1)$ is chosen to denote the signal V$_{PD,A}$ during $t_1$ time interval and also applicable to similar terms.  {\color{black}In the next stage the output of AMP I is chopped using CH II, which demodulates the photodiode signal to the baseband and modulate the offset and low-frequency noise to $f_{ch}$. CH II consists of switches S$_5$-S$_8$ operated on same non-overlapping clocks $\phi$ and ${\phi}$'. The differential output of the CH II is amplified by single-ended difference amplifier AMP II. The output of AMP II is fed back to the Pixel$_{A}$ and Pixel$_{B}$ to close the loop. Thus, AMP I and II both are required to form a closed loop unity gain system.   
    
If AMP I and AMP II has a voltage gain of $A_1$ and $A_2$, offset of $V_{of1}$ and $V_{of2}$, low-frequency noise of $N_{Am1}$ and $N_{Am2}$, respectively, the output signal V$_{out}$ is expressed as}
\begin{equation}
\label{vouta}
\centering
\begin{aligned}
V_{out} = [V_{PD,A}(t_1)] + V_{PD,B}(t_2)] 
+ [N_{sf,A}(t_1) - N_{sf,A}(t_2)] \\ - [N_{sf,B}(t_1) - N_{sf,B}(t_2)] 
+ \frac{N_{Am1}(t_1) - N_{Am1}(t_2)}{A_1} \\ + \frac{V_{of1}(t_1) - V_{of1}(t_2)}{A_1} 
+ \frac{N_2 + V_{off2}}{2A_1A_2}. 
\end{aligned}
\end{equation}

    \begin{figure*} 
    \centering
    \includegraphics[scale =0.6]{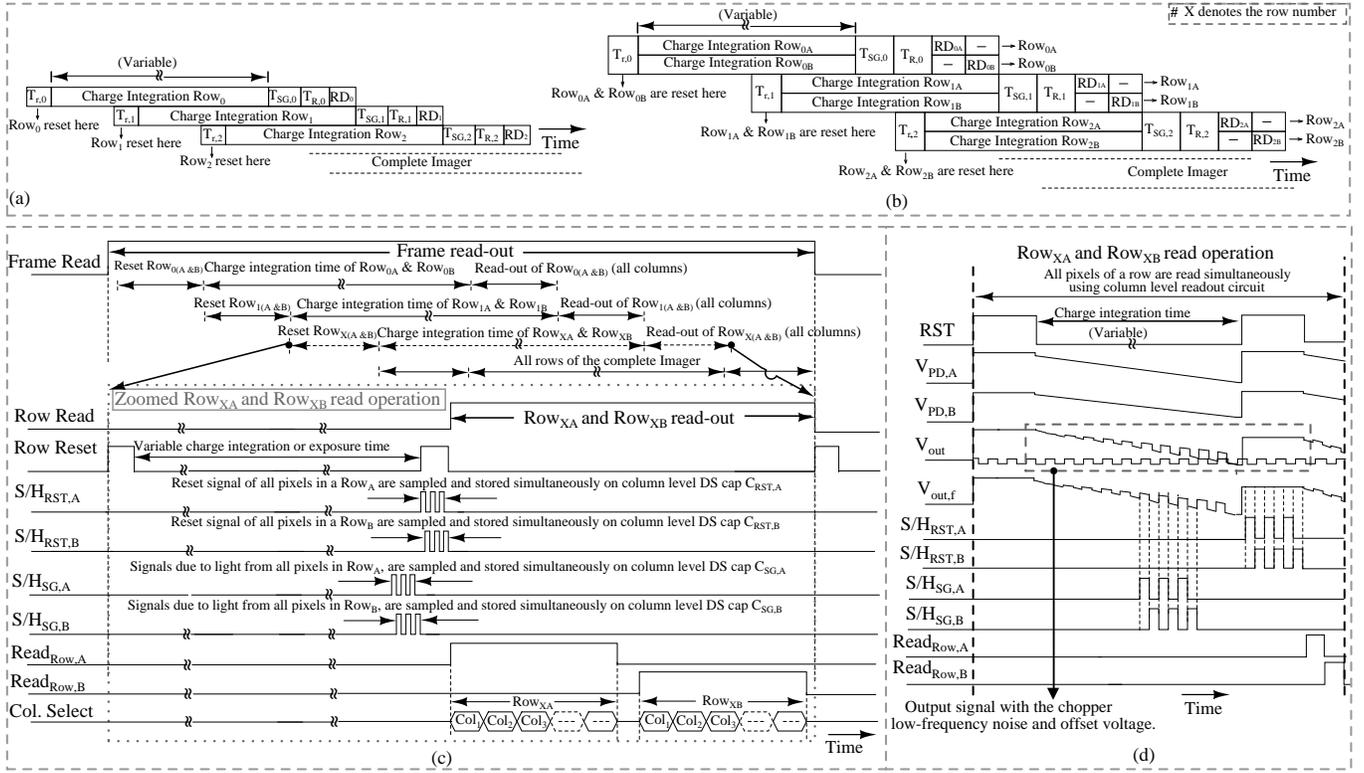}
    \caption {{\color{black} (a) Conventional rolling shutter (b) proposed readout. Pixels of Row$_{XA}$ and Row$_{XB}$ are reset during the time interval T$_{r,X}$. After a variable charge integration time, the output signal of all pixels of Row$_{XA}$ and Row$_{XB}$ are sampled and stored on column level capacitors during T$_{SG,X}$. Row$_{XA}$ and Row$_{XB}$ are reset again and sampled and stored on column level capacitors during T$_{R,X}$. Row$_{XA}$ and Row$_{XB}$ are read-out during RD$_{XA}$ and RD$_{XB}$, respectively, (c) Timing diagram of a frame read-out for the imager based on proposed technique, and (d) Timing diagram of Row$_{XA}$ and  Row$_{XB}$ read-out.}}
    \label{TIMING} 
    \end{figure*} 
 
If the small signal voltage gain values $A_1$ and $A_2$ are very high, then only the photodiode signals V$_{PD,A}$ and V$_{PD,B}$ along with 1$/f$ noise from the source followers dominate and the output signal  V$_{out}$ can be simplified as 	 
\begin{equation}
\label{voutb}
\begin{aligned}
 V_{out} \approx [V_{PD,A}(t_1) + V_{PD,B}(t_2)] ~~~~~~~~~~~
\\ +~[N_{sf,A}(t_1) - N_{sf,A}(t_2)]- [N_{sf,B}(t_1) - N_{sf,B}(t_2)]. 
\end{aligned}
\end{equation}


{\color{black} In (2) and (3), it is assumed that the chopping frequency is much higher than the low-frequency noise corner frequency and the noise pairs like N$_{sf,A}$($t_1$) and N$_{sf,A}$($t_2$) are correlated due to high $f_{ch}$~\cite{paper:SWITCH_Gierkink99,paper:Mypaper2}.
%
%
}
{\color{black}
 To suppress the overall input referred noise and offset at the output (from the amplifier stages), a two-stage high gain amplifier is used. The amplifier unity gain-bandwidth is 42 MHz with a phase margin is greater than 65$^0$ and with a maximum power consumption of {526 $\mu$W (160 $\mu$A bias current and 3.3 V supply voltage)}. The first stage of the opamp is a differential input/differential output folded cascode amplifier (AMP I with small signal volt. gain of 65 dB), which is followed by a difference amplifier with a single-ended output (AMP II with small signal volt. gain of 40 dB) to achieve an overall gain of 105 dB. }         
%
%
%
%
%
%
%
  
The output signal V$_{out}$ is continuous and composed of Pixel$_{A}$ output for time duration $t_1$ and Pixel$_{B}$ output for time duration $t_2$, periodically. The switches used for chopping introduces ripples at the output. These ripples are generated due to clock feed-through of the overlapping capacitance present between the drain and gate of the switching transistors. A switched capacitor low-pass filter is used to block the ripples present in the output signal \cite{paper:Filter1}. 


As the dynamic range (DR) of an active pixel sensor is defined as the ratio between the saturation and random noise floor, thus, can be given as:
\begin{equation}
\text{DR} = 10 \cdot log\bigg(\frac{N_{sat}}{n_{drn}}\bigg),
\end{equation} 
where $N_{sat}$ is the saturation level signal of the pixel and $n_{drn}$ is the pixel dark random noise. As the proposed technique reduces the random noise, thus, it also enhances the DR of the image sensor. The photodiode signal {\color{black}gets} buffered through a chopper amplifier including SF, high gain amplifier stage I and II (configured in closed loop with unity gain) and the final output is fed back to one of the inputs of first chopper CH I. Hence, the continuous output of the closed-loop chopper amplifier is virtually short with the photodiode output node. The high gain of the amplifier (105 dB) makes the output follow the photodiode node linearly for a wide range of light integration, increasing the output swing and dynamic range.
%
%
%
{\color{black}
\subsection{Imager Read-out Operation}
\label{READ_OUT}
The architecture of the image sensor using the proposed technique is shown in Fig. \ref{fig:ARCH}. As the in-pixel chopping is applied to the conventional 3T pixel, the read-out is, therefore, very similar to a 3T pixel architecture which is progressive in nature. In the proposed technique the read-out is based on conventional rolling shutter mode. The conventional and proposed read-out mode is shown in Fig. \ref{TIMING}(b) and (c),  respectively. 
However, in the proposed architecture instead of a single row, two adjacent rows, Row$ _{XA} $ and Row$ _{XB} $ (X is used to denote the row number, for example, Row$_{1A}$ and Row$_{1B}$) are selected together for readout.
Charge integration on photodiode, charge to voltage conversion, chopping/de-chopping of the photodiode signal, signal due to light/reset level sample and hold, double sampling (DS) and low-frequency noise filtering are carried out on the pixel pairs (i.e. Pixel$_{ 0A }$-Pixel$_{ 0B }$, Pixel$_{ 1A }$-Pixel$_{ 1B }$, Pixel$_{ 2A }$-Pixel$_{ 2B }$.....) for Row$_{XA }$ and Row$_{XB}$  together. These operations on each pair of the pixels of the selected rows (Row$ _{XA} $ and Row$ _{XB} $) are carried out in parallel using column level circuit. The timing diagram of the proposed in-pixel chopping architecture, as shown in Fig. \ref{TIMING}(d) and (e).   \par
The double sampling circuit is modified to sample and hold the reset and signal of the pixel pair of adjacent rows, as shown in Fig. \ref{Ckt_inPxl_Chop}. During the reset phase Row$_{XA}$ and Row$_{XB}$ are reset and after a variable charge integration or exposure time the signals from Pixel$_A$ and Pixel$_B$ are sampled on sampling capacitors  C$_{SG,A}$ and C$_{SG,B}$, respectively.
{\color{black} Then, Row$_{XA}$ and Row$_{XB}$ are reset again and the reset levels of Pixel$_A$ and Pixel$_B$ are sampled on the reset capacitors C$_{RST,A}$ and C$_{RST,B}$. }
Switch S/H$ _{RST,A} $ and S/H$ _{RST,B} $ are ON for repetitive and non-overlapping time intervals t$_{ 1 }$ and t$_{ 2 }$, respectively,  sampling the reset levels, while, switch S/H$ _{SG,A} $ and S/H$ _{SG,B} $ are ON similarly, sampling the output signals.
This sampling is performed on all pixel pairs of the Row$_A$ and Row$_B$, simultaneously using column level sample and hold circuit.
After storage of the reset level and output signals, delta differential sampling (DDS) is performed to cancel the pixel level fixed pattern noise (FPN).
\par
However, during the double sampling, the sampled $k$T/C noise  components of the reset switch are non-correlated as they come from two different reset phases, Thus, instead of elimination, DDS eventually increases the thermal noise, which is a well-known limitation of 3T pixel readout.}
\begin{figure*}
	\includegraphics[scale = 0.52]{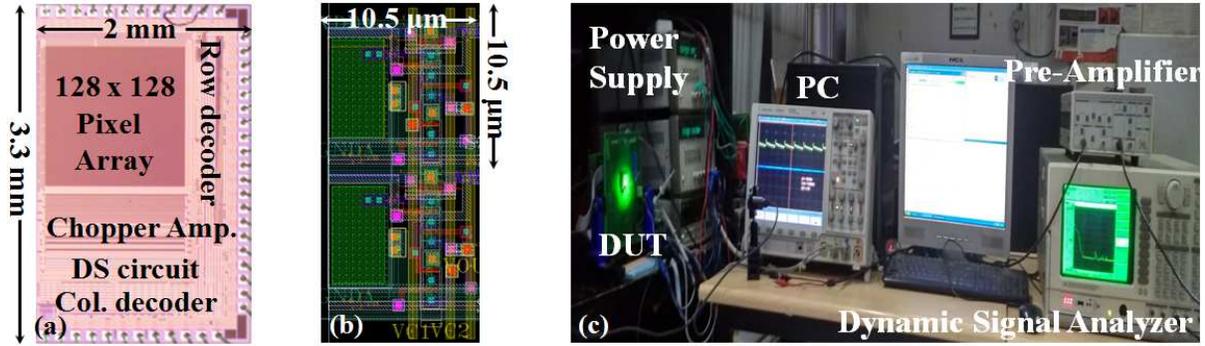} 
	\centering 
	\caption{{\color{black}(a) Chip micro-photograph, (b) 2$\times$1 Pixel array layout, and (c) Measurement setup.}}
	\label{DUT}
\end{figure*} 
\begin{figure*}
\includegraphics[scale = 0.43]{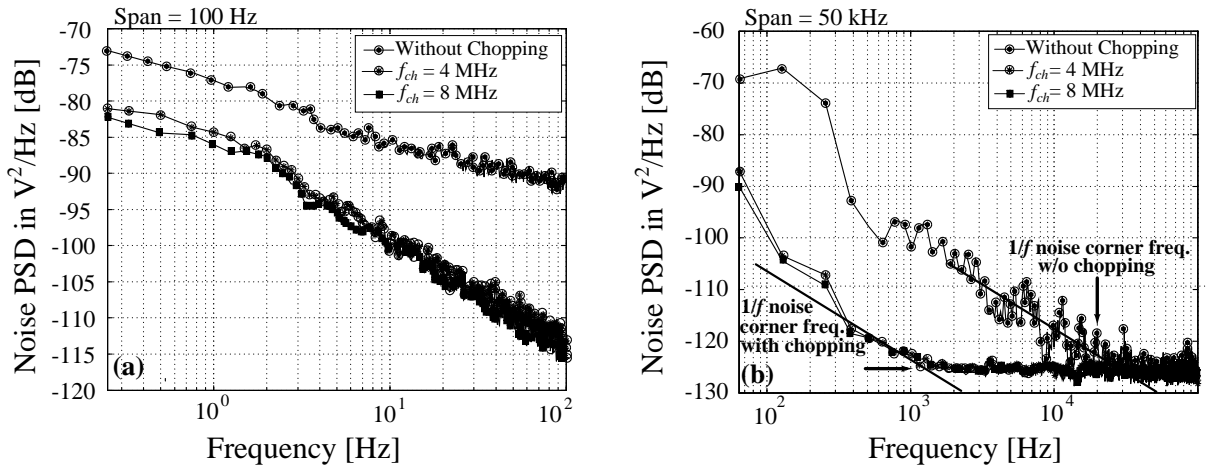} 
\centering 
 \caption{{\color{black} Measured noise PSD at the output of the DS circuit without and with in-pixel chopping (at $f_{ch}$ = 4 MHz and 8 MHz) for sampling frequency span of (a) 100 Hz, (b) 50 kHz.}}
\label{MSR_NOISE}
\end{figure*}
\begin{figure*}
\includegraphics[scale = 0.295]{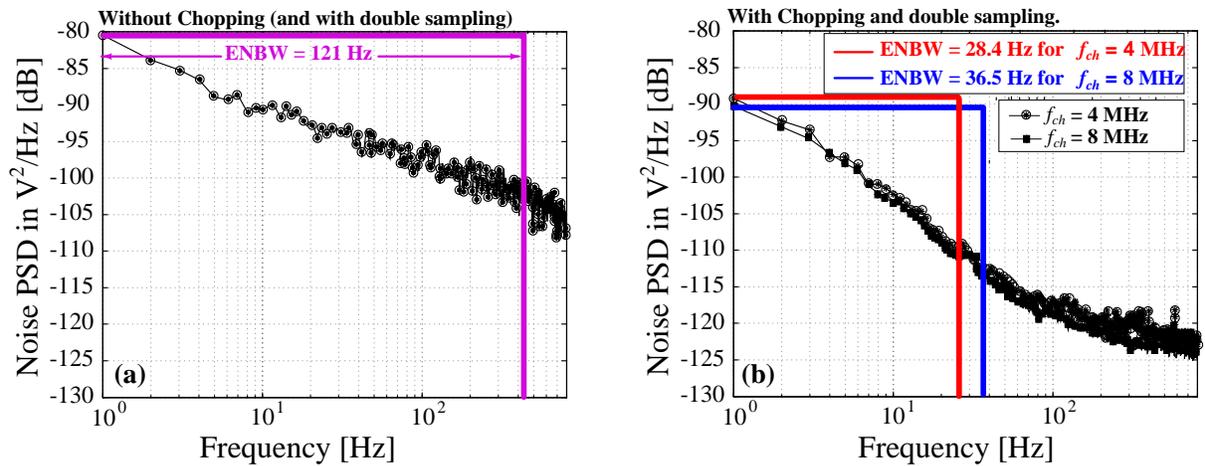} 
\centering 
 \caption{{\color{black} Measured noise PSD at the output of the DS circuit without and with in-pixel chopping (at $f_{ch}$ = 4 MHz and 8 MHz) for sampling frequency span of 800 Hz; Reduction in equivalent noise bandwidth (ENBW) and 1/$f$ noise corner frequency with chopping is also shown in figures (a) and (b), respectively.}}
\label{MSR_NOISE2}
\end{figure*}
\begin{figure*}[t]
\includegraphics[scale = 0.5]{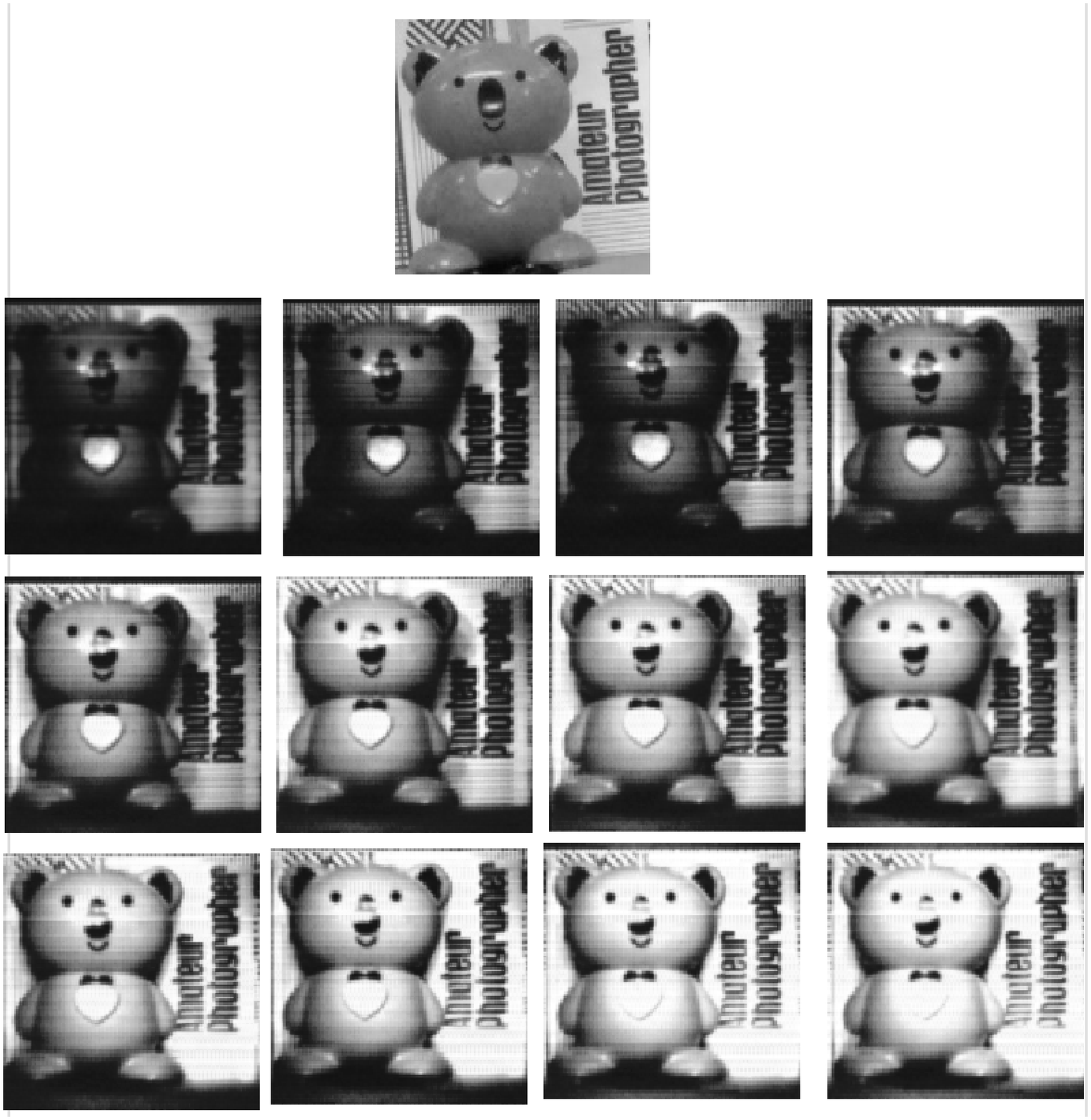} 
\centering 
 \caption{{\color{black} Test images taken at a variable integration time from 100 $\mu$s to 2 ms  and a fixed illumination of around 300 lux to show the response of the imager.}}
\label{Images}
\end{figure*}

 \section{Low-Frequency Noise Reduction using Chopper Stabilization Technique}
 \label{sec:CHOP_ANALISYS}
 The chopper stabilization technique was introduced to implement ac coupled amplifier with high precision low-frequency gain. The chopping technique does not involve the sampling of the input signal, but rather it up-convert the signal to the chopping frequency and then down-convert it back to the baseband.
 In the chopper stabilization technique the continuous time input signal V$_{\text{in}}$, is multiplied with \textit{m(t)}, a train of square-wave pulses with the fundamental frequency of $f_\text{{ch}}$ = 1/T${_p}$, where T${_p}$ is the time period of \textit{m(t)}. The input-referred noise of the source follower followed by the amplifier (without chopping) can be given as \cite{paper:Enz_CDS_chop}: 
 \begin{equation}
 S_{\text{N}}(f)^2 = S_{\text{0}} \cdot \bigg(1 + \frac{f_c}{|f|}\bigg) \cdot \frac{1}{\bigg[1+ \bigg(\frac{f}{f_0}\bigg)^2\bigg]},
 \label{inref_noise_psd_wo_chop}  
 \end{equation}    
 where $S_{\text{0}}$ denotes the white noise component, $f_0$ is the cut-off frequency or the dominant pole of the amplifier, and $f_c$ represents the corner frequency of the $1/f$ noise, which is the value of frequency at which the flicker noise PSD becomes equal to the thermal or white noise. In this technique, the input signal is first chopped alone and gets modulated to the odd harmonics of the signal $m(t)$. After processed by SF and amplifier, it gets chopped again and demodulated to the base-band. In this process, the noise of the SF and amplifier is chopped only once and get transposed to the to the odd harmonics of the signal $m(t)$. Thus, the chopped noise PSD can be given as [\cite{paper:Enz_CDS_chop}],
 \begin{equation}
 S_{\text{CN}}(f) = \bigg(\frac{2}{\pi}\bigg)^2  \sum_{n (odd) = -\infty }^{\infty} \frac{1}{n^2} \cdot S_\text{N}(f) \cdot \bigg(f - \frac{n}{T}\bigg),\\
 \label{inref_chopped_noise}  
 \end{equation}  
 If the chopping frequency of much higher than the limit of base-band and also the cut-off frequency of the amplifier ($f_0$) is much higher than the chopping frequency, the white noise component of the chopped noise PSD (with the Nyquist criteria $2|f| > f_{ch}$) can be approximated as :

 \begin{equation}
 S_{\text{CN-white}}(f) =  S_{\text{CN-white}}(f = 0)
 = S_{\text{0}} \cdot   
 \bigg[1- 
 \frac{\text{tanh} \bigg( \frac{\pi}{2} f_0 T_p \bigg)   }{\frac{\pi}{2} f_0 T_p} \bigg]
 \label{white_noise}  
 \end{equation}    
 If the cut-off frequency of the amplifier ($f_0$) is also much higher than the chopping frequency, thus, for $f_0 T_p>>1$, the expression can further be approximated as:
 \begin{equation}
 S_{\text{CN-white}}(f) \approx S_{\text{0}}\\
 \label{white_noise_simp}  
 \end{equation}    
 For the low-frequency noise reduction using auto-zeroing which, involves in the sampling of the input signal and introduces aliasing of the signal in-band white noise. Hence, baseband noise PSD increases with the increase in the ratio of noise BW and sampling frequency. However, (\ref{inref_chopped_noise}) indicates that the chopping technique does not increase the white noise at all and for  $f_0 T_p>>1$ the white noise PSD tends to the value of original white noise without chopping.  
 \par
 The effect of chopping on low-frequency noise and systematic offset of the amplifier can be analyzed as follows: For $f_0>>f_{\text{ch}}$, the input noise PSD of chopper can be approximated as \cite{paper:Enz_CDS_chop}:
 \begin{equation}
 S_{N-\text{1/\textit{f}}}(f) = S_{\text{0}} \cdot \bigg(\frac{f_c}{|f|}\bigg) =  \frac{f_c T_p}{|f T_p|},
 \label{lf_chppednoise}  
 \end{equation} 
 With the above approximation it can be analyzed that the low-frequency pole from the $1/f$ noise component of (\ref{inref_chopped_noise}), is shifted to $f_{\text{ch}}$ and to its odd harmonics. Further, the chopped $1/f$ noise PSD inside the signal frequency band can be approximated as:
  \begin{equation}
 S_{\text{CN-1/f}}(f) \approx K S_0 f_c T_p.\\
 \label{chopped_1/f_noise_appx}  
 \end{equation}  
 where $K$ is approximately 0.852 for above-mentioned approximations. Thus, the resultant total input referred baseband chopped noise can be given as:  
 \begin{equation}
 S_{\text{CN}}(f) \approx S_0(1 + Kf_c T_p). ~\newline \text{for} ~|fT_p| \leq 0.5~\text{and}~ f_0 T_p>>1.
 \\
 \label{Tot_Appx_chopped_noise}  
 \end{equation}  
 Thus, the resultant baseband noise PSD after chopping is approximately the white noise PSD if chopping frequency is chosen as much higher then the corner frequency.    

\section{Sensor Overview and Measurement Results}
\label{MSR_SIM}
{\color{black} The prototype image sensor consists of an array of 128$\times$128 modified 3T-pixels with chopping, column level read-out circuitry including a chopper amplifier with low-pass filter and a modified double sampling circuit, chip level row and column decoders, and input/output buffers. The sensor is fabricated in AMS 0.35 \textbf{$\mu$}m CMOS OPTO process. The micro-photograph of the sensor and the layout of a pixel pair are shown in Fig. \ref{DUT}(a) and (b), respectively. The modified 3T pixel with a chopper inside is laid out at the pitch of 10.5 $\mu$m (horizontal and vertical both) with 30$\%$ fill-factor (FF). For sensing the ambient light signal an n-well/p-sub photodiode is used, which is cost effective in terms of fabrication, as compared to a pinned-photodiode (PPD). {\color{black}  All transistors inside the pixel and choppers are nMOS transistors with minimum dimensions (width of 0.4 $\mu$m and length of 0.35 $\mu$m) to maintain the fill-factor.
		
		To test the complete imager, an off-chip 14-bit analog-to-digital converter (AD9822) with a 2-volt output swing, is used to convert an analog output into a digital output. Out of 14 bits, first 12 bits from MSB, are used for measurements. A frame-grabber (NI PCI-1424) and a PC are used for processing of the digital data. All external control signals/clocks for all the on-chip and off-chip circuitry are generated using an FPGA (Altera EPIC3T144C8).
		
		\subsection{Low-Frequency Noise Measurement}
		\label{NOISE_MSR}
		
		The measurement setup for the low-frequency noise is shown in {\color{black}Fig. \ref{DUT}(c), which is similar to that reported in  
			\cite{paper:wei01,paper:Mypaper2,paper:SWITCH_VEN_DER_2000}. The low noise voltage preamplifier (SR560) is used to amplify the noise power of the DUT. The input referred noise of the voltage preamplifier is as low as 4~nV/$\sqrt{\text{Hz}}$, which makes it quite suitable for precise noise measurement. After amplified by the voltage preamplifier, the noise voltage is analyzed using Dynamic Signal Analyzer (DSA - SR785). The DSA plots the  Fourier transform of the noise voltage signal from the pre-amplifier. The input referred noise of the SR785 inputs is about 10~nV$_{\text{rms}}/\sqrt{\text{Hz}}$. The input referred noise of the analog-to-digital converter (ADC) in SR785 is about 300~nV$_{\text{rms}}/\sqrt{\text{Hz}}$ (referenced to a full scale of 1 V$_{\text{pk}}$)}

		If the thermal noise components from reset switch come from the same reset phase, they can be eliminated from the output by subtraction using double sampling circuit. The double sampling circuit acts as a high-pass filter for the low-frequency noise. Thus, along with the thermal noise, it can also reduce the fixed pattern noise (FPN) and low-frequency noise. However, the reduction in the low-frequency noise power depends on the sample-to-sample time period ($t_\text{s}$) \cite{TED16}, \cite{paper:CHOP4}. The transfer function of the double sampling is given by \cite{paper:James_Jen_CCD}:

\begin{equation}
|H_{\text{CDS}}(f)|^2 = 2 - 2\cdot cos({2\pi f t_s}).
\end{equation}
A higher rejection in the low-frequency noise can be achieved by keeping $t_s$ smaller because of correlation between sampled noise components. For $t_s = 0$ the all sampled noise components are correlated and thus, the subtraction results in complete elimination of the noise. While, for $t_s = \infty$, the resultant noise is equal to the quadrature sum of the sampled noise components, for either uncorrelated noise samples are added or subtracted.

The present technique does not have this limitation and to prove its effectiveness, the noise PSDs have been measured at the output of the pixel after processed by DS circuit with and without chopping, and then the comparison of the measured noise power is reported. During measurements, the sample-to-sample time period is kept as 32 $\mu$s, and both signal and reset levels are sampled in a pulse-width of 16 $\mu$s. Thus, each sampling pulse includes a total of 64 and 128 chopping pulses for $f_{ch}$ = 4 MHz and 8 MHz, respectively. The measured noise PSD curves for varying $f_{ch}$ in shorter sampling frequency span of 100 Hz, 800 Hz, and in a longer span of 50 kHz, are shown in {\color{black}Fig. \ref{MSR_NOISE} (a), (b), and (c)}, respectively. For each span, the noise PSD of the pixel after double sampling is shown, without chopping and with in-pixel chopping for $f_{ch}$ equal to 4 MHz and 8 MHz. To improve the accuracy, each noise PSD curve is plotted after taking an RMS average of 1000 measured samples. }
%
%
%
%

\begin{figure}
	\def\putbox#1#2#3#4{\makebox[0in][l]{\makebox[#1][l]{}\raisebox{\baselineskip}[0in][0in]{\raisebox{#2}[0in][0in]{\scalebox{#3}{#4}}}}}
	\def\rightbox#1{\makebox[0in][r]{#1}}
	\def\centbox#1{\makebox[0in]{#1}}
	\def\topbox#1{\raisebox{-0.60\baselineskip}[0in][0in]{#1}}
	\def\midbox#1{\raisebox{-0.20\baselineskip}[0in][0in]{#1}}
		\scalebox{0.3426}{
		\normalsize
			\hspace{1.5 in}
			\parbox{1 in}{
			\includegraphics[scale=2]{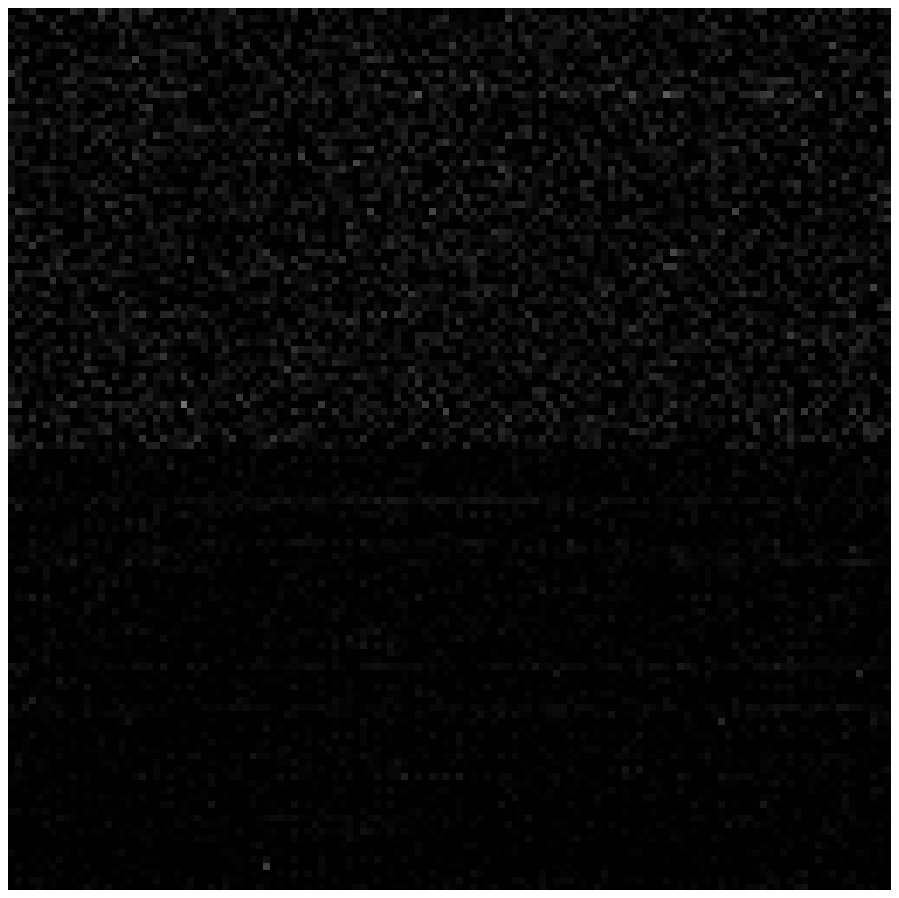}\\
			\putbox{0.2 in}{6.5in}{3}{{\color{white} (a) without chopping}}%
			\putbox{0.2 in}{2.9in}{3}{{\color{white} (b) with chopping}}%
		} 
	} 
	\caption{{\color{black} Test images under no ambiance light (a) without and (b) with in-pixel chopping.}}
		\label{DARK_IMG}                                                                  
\end{figure}

{\color{black}  As shown in the Fig. \ref{MSR_NOISE} (b) the 1/$f$ corner frequency which is around 20 kHz without chopping (with DS), shifts to around 1.1 kHz with chopping and DS. The integrated noise power from 1 Hz to 20 kHz (thermal noise floor reaches around 20 kHz) at the output after DS without chopping is -60.15 dB (noise in volts = 982.3 $\mu$V$_{RMS}$). With DS and chopping the integrated noise power reduces to -73.47 dB (212 $\mu$V$_{RMS}$) and -74.37 dB (191.2 $\mu$V$_{RMS}$) for $f_{ch}$= 4 MHz and 8 MHz, respectively, which shows the noise power reduction of 13.31 dB and 14.22 dB, respectively.} The spot noise power at 1 Hz sampling frequency without chopping (with DS) is -81 dB (7.94 nV$^2_{RMS}$/Hz) and with chopping and DS is -88 dB (1.58 nV$^2_{RMS}$/Hz) and -90.1 dB (0.977 nV$^2_{RMS}$/Hz) for $f_{ch}=$ 4 MHz and 8 MHz, respectively, as shown in Fig. \ref{MSR_NOISE} (a). For the total integrated noise power in the frequency band from 1 Hz to 20 kHz, the equivalent noise bandwidth (ENBW) is 121 Hz with DS and without chopping, which reduces to 28.35 Hz for $f_{ch}=$ 4 MHz and to 36.5 Hz for $f_{ch}=$ 8 MHz with chopping and DS, which is shown in the Fig. \ref{MSR_NOISE} (a).

\begin{table}
\centering
\caption{Performance characteristics of the reported CMOS Image sensor}
\label{tab:Table 2}
\begin{tabular}{p{3.7 cm}|p{4.2 cm} }
            
 \textbf{Parameter}            &  \textbf{Performance}                \\ \hline \hline
 Technology                    &  0.35 $\mu$m AMS 2P4M  CMOS OPTO     \\ \hline
 Power supply                  &  3.3 V (analog/digital)              \\ \hline
 Chip size                     &  3.3 mm x 2.0 mm                     \\ \hline
 Array size                    &  128$\times$128                      \\ \hline
 Photo-detector                &  n-well/p-sub photodiode             \\ \hline
 Pixel architecture            &  Modified 3T APS					  \\ \hline
 Transistors per pixel         &  4									  \\ \hline
 Pixel pitch                   &  10.5 $\mu$m (V) \& 10.5 $\mu$m (H)  \\ \hline
 Fill-Factor 			       &  29.5 \%                             \\ \hline
 Full Well Capacity	           &  88 ke$^-$                           \\ \hline
 Lens (used for focusing)      &  F$\#$1.4/16 mm focal length         \\ \hline
 Frame-rate                    &  30 fps                              \\ \hline
 Chopping frequency ($f_{ch}$) &  4 MHz and 8 MHz                     \\ \hline
 Fixed pattern noise           &  4 to 5\% 				    		  \\ \hline
 Sensitivity                   &  2.5  V/lx-s                         \\ \hline
 Output signal swing           &  1.9 V                               \\ \hline
 Conversion gain               &  19.75 $\mu$V/e$^-$                  \\ \hline 
 Random noise                  &  280 $\mu$V$_{\text{RMS}}$ ($f_{\text{ch}}$ = 8 MHz) \                         \\ \hline
 Dynamic range (w/o chopping)  &  65 dB                        \\ \hline 
 Dynamic range (with chopping) &  76 dB                        \\ \hline
 ADC resolution (off-chip)     &  12-bit                              \\ \hline
 Saturation level              &  3800 DN                           \\ \hline
 Dark Current                  &  930 $\mu$V/s                            \\ \hline
  \hline
  \end{tabular}
 \label{TAB:Table1}
 \end{table}

\begin{table*}
\normalfont                   
\centering
\caption{Performance Comparison of Recently Reported CMOS Image Sensors with Noise Reduction Techniques.}
\label{tab:Table 2}
\begin{tabular}{M{3.7 cm}|M{2.3cm}|M{2.3cm}|M{2.3cm}|M{2.5cm}|M{2.2cm}}
\hline

 \textbf{Noise Reduction Technique}     
& {pMOS Common Source Pixel Amplifier}
& {Burried Channel nMOS SF, Multiple Sampling with SSADC}
& {Thin Oxide pMOS SF}
& {Correlated Multiple sampling (CMS)}
& {{In-pixel Chopping }}
\\   \hline
     \hline
\textbf{Reference and Year}
& ISSCC\cite{ISSCC2011} - 2011   
& ISSCC \cite{ISSCC2012} - 2012  
& TED \cite{TED16} - 2016 
& Sensors J. \cite{CMS_SJ2012} - 2012  
& This work                            \\ \hline 

\textbf{Technology}             
& 180 nm - CIS   
& 180 nm - CIS  
& 180 nm - CIS   
& 180 nm - CIS  
& 350 nm - CMOS                        \\ \hline 

\textbf{Array size}
& 256$\times$256
& 128$\times$196  
& -NA-  
& 7$\times$2  
& 128$\times$128                              \\ \hline

\textbf{Pixel pitch [$\mu$m]}
& 11  
& 10  
& 7.5  
& 10  
& 10.5                                        \\ \hline

\textbf{Fill-Factor [$\%$]}
& 50  
& -NA-  
& 66  
& -NA-  
& 30                                          \\ \hline

\textbf{Conversion Gain [$\mu$V$_\text{RMS}$/e$^-$]}
& 300  
& 45  
& 185  
& 45  
& 19.5                                        \\ \hline

\textbf{Pixel Architecture     }
& 4T APS 
& 4T APS 
& 4T APS  
& 4T APS 
& Modified 3T APS                             \\ \hline

\textbf{Photo detector}
& PPD  
& PPD  
& PPD   
& PPD  
& n-well/p-sub PD                     \\ \hline

\textbf{Temporal noise [$\mu$V$_\text{RMS}$]}
& 258   
& 31.5  
& 74  
& 127  
& 280                             \\ \hline

\textbf{Dynamic Range [dB]     }
& 90  
& -NA-  
& -NA-  
& -NA-  
& 76   					               \\  \hline   \hline  
                            
\end{tabular}
\end{table*}

\subsection{Statistically Measured Temporal Noise using PTC Method}
\label{NOISE_MSR}
The dark random noise of the proposed image sensor is also calculated statistically using the photon-transfer-curve method \cite{paper:James_Jen_CCD}. The analog output of the pixel after double sampling is converted into the digital number (DN) by using the upper 12-bits of a 14-bit off-chip ADC. To obtain the PTC of the image sensor, the standard deviation (SD) is plotted against the output signal from 128$\times$128 pixels. For PTC calculation, an average of 100 frames has been taken before calculating the SD of pixel outputs.

To calculate the dark random noise, the sensor was kept in dark room, exposed with no external light input, at normal room temperature, and without any optical filter. The photon-transfer-curve (PTC) is calculated by the variance of the measured output of 128$\times$128 pixels, which are uniformly illuminated, with an LED source with a variable voltage source, using an integrating sphere. For remaining part of the PTC each pixel of the image was illuminated with constant uniform light of 650-lux, and to obtain the variable output signal, the integration time was made variable ranging from 50 $\mu$s to 2 ms. The variation of the pixel output (after double sampling) with respect to the integration time at a given intensity level is shown in Fig. \ref{Sens}. The sensitivity is calculated as 2.5 V/lx-s, from the slope of the curve. The gain of the SF with chopper amplifier configured in unity feed-back is almost 1. During measurements, internal gain of the ADC was also kept as unity.       

 \begin{figure*}[ht!]
 	\includegraphics[scale = 0.9]{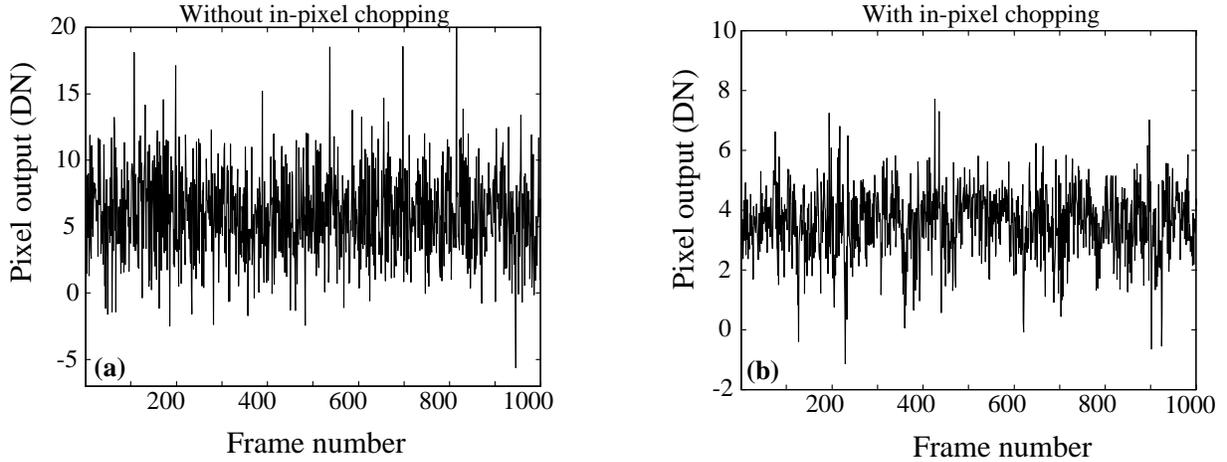} 
 	\centering 
 	\caption{{\color{black} The digital output (in digital number or DN) of each pixel in a frame of 128$\times$128 pixels after double sampling in a frame (average of 100 frames), (after subtracting the dark frame) to show the statistical variation (a) without chopping, (b) with both chopping.}}
 	\label{VAR}
 \end{figure*}
  
To calculate the random noise, the standard deviation of an averaged frame under no light condition is measured without and with the in-pixel chopping. The fixed pattern noise or offset correction is done by the dark frame subtraction.
To show the reduced variance, the output of pixels (after double sampling) is without and with in-pixel chopping is plotted in Fig. \ref{VAR} (a) and (b), respectively. The resultant standard deviation (random noise) of the output data without chopping (with DS) is 2.1 LSB, which reduces to 0.6 LSB with chopping and DS.
{\color{black}The test image of a 128$\times$128 pixel array (after dark frame subtraction) is shown in Fig. \ref{DARK_IMG}}. The upper 64 and lower 64 rows ($\times$ 128 columns) are showing the output of pixels (after double sampling) without and with in-pixel chopping, respectively. The reduction in the variation in the output can easily be concluded from the test image shown in Fig. \ref{DARK_IMG}. The output in DN of all the pixels is plotted in Fig. \ref{VAR} and it is evident from the comparison that the variation in the output after double sampling is lesser among the pixels with in-pixel chopping as compared to the case without chopping.  
}
\begin{figure}[t]
\includegraphics[scale = 0.42]{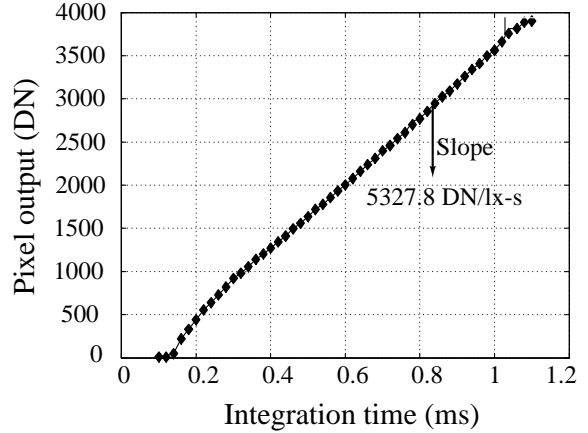} 
\centering 
 \caption{{\color{black} The photo-response of the APS with in-pixel chopping under variable integration time and fixed uniform illumination of 650 lux.}}
\label{Sens}
\end{figure}
\subsection{Dynamic Range}
\label{DR}
The dynamic range is also calculated using the PTC method. As mentioned to obtained the PTC, the sensor exposed uniformly with the light input provided by an LED and intensity was kept constant at 650 lux, while integration time was made variable using the clocks generated by an FPGA. The variance and mean value are extracted from an average of 100 images for each integration time.
The resultant standard deviation (random noise) in the dark of the output data without chopping (with DS) is 2.1 DN, which reduces to 0.6 DN with chopping and DS. While, for both the cases the variance starts decreasing around mean output value of around 3800 DN or approximately 1.85 V, due to the readout circuit saturation. Thus, the resultant DR, using \ref{DR}, is calculated as 65 dB without chopping (and with double sampling) which, enhances to 76 dB by using chopping along with double sampling.
To demonstrate the dynamic range test images, shown in Fig. \ref{Images}, taken from the prototype image sensor under fixed light exposure and variable integration time from 100 $\mu$s to 1.2 ms. \par
By estimating the slope factor in the linear part of the PTC is used to calculate the conversion of the image sensor which comes around 19.5 $\mu$V/e$^-$, with the unity gain of the readout chain. As the saturation level of the PTC curve arrives around 1.85 V, the full well capacity (FWC) is about 88 ke$^-$.

\subsection{Dark Current}
\label{Dark_I}
To measure the dark current, the prototype sensor was kept in dark chamber, covered with a flap, to make sure no incident light. The measurement is performed at room temperature around 27$^o$C, with variable integration time. The read-out chain gain was kept as unity. The output signal of each pixel of the image sensor is measured for the integration time varying from 1 ms to 100 s. A slope of the linear region of the curve between the output and integration time of each pixel is measured and an average of all slopes is taken.
A value of the averaged slope reflects the dark current of the image sensor, which is calculated as 930 $\mu$V/s. As the integration time is very small during imaging applications, this value of the dark current is quite negligible as compared to the measured read noise floor.

\section{Discussion And Conclusion}
\label{CONCLUSION}
The chopper stabilization reduces the noise by modulation and without aliasing of the white noise. The in-pixel chopping is realized using an additional switch in to a conventional 3T pixel and used to reduce the low-frequency noise of the source follower. Using the technique a reduction of {11} dB, in the integrated noise power, is obtained. The measured parameters of the proposed imager are summarized in the table \ref{TAB:Table1} while, the performance comparison of this work with other recently proposed imagers is shown in table \ref{tab:Table 2}. As shown in the table \ref{tab:Table 2}, the input-referred temporal noise in the proposed work is quite comparable to the other image sensors. The imager in this work is employed with an n-well/p-sub photodiode which is inherently more noisy but rather more cheap in the cost, as compared to the pinned photodiode (PPD). The proposed technique reduces the input referred temporal noise from 1.1 mV$_\text{RMS}$ to 280 $\mu$V$_\text{RMS}$. It is noted that the reported noise is also containing the thermal noise, low-frequency noise, and quantization noise (LSB/$\sqrt{12}$) components of the external ADC. In addition, this technique is applied to the 3T pixel and the conventional 3T pixel read-out increases the thermal noise rather than cancel it. However, in a 4T pixel with a pinned photodiode a true CDS can be implemented to cancel the thermal noise component of the reset switch. Thus, it can be claimed that using the proposed technique a comparable performance has been achieved with a lower cost technology with the proposed sensor.      
\par
By obtaining a reduced temporal noise,  the dynamic range of the image sensor also enhanced from 65 dB to 76 dB, which eventually improves the quality of the image. As the noise reduction using the proposed technique does not depend on the size of the transistor, a minimum sized SF is used in the APS as a voltage buffer. Choice of an SF with minimum area partially compensate for the increase in the fill-factor due to an additional in-pixel switch. And the low-frequency noise of the minimum sized SF is reduced using in-pixel chopping. } } 
To implement the chopper stabilization, the additional circuitry which includes a high gain chopper amplifier and a low-pass filter is required in each column. These extra column level blocks does not affect the pixel fill-factor but rather, increases the overall power consumption of the image sensor. A lower power consumption can be achieved by compromising the bandwidth of the amplifier, however, it will restrict the maximum value of the chopping frequency.\par
In this work, the proposed in-pixel chopping technique is implemented in a conventional 3T APS, while further investigation can also make it enable to implement the in-pixel chopping inside a 4T active pixel sensor. 

\section{Acknowledgment}
\label{ACK}
The authors would like to wish thank to Dr. Shauri Chettarjee and Dr. Gajendranath
Chowdhary for providing their expertise in the circuit designing. The authors would also like to thank Dr. Kushal Shah for providing his valuable inputs to understand the modeling of the low-frequency noise.
The authors also thank Mr. R. K. Singh for providing his valuable technical inputs and help in designing of the test boards for sensor measurement. 


\end{document}